# Exfoliation of layered Na-ion anode material Na$_2$Ti$_3$O$_7$ for enhanced capacity and cyclability

Maria A. Tsiamtsouri [ab], Phoebe K. Allan [acd], Andrew J. Pell [a]†, Joshua M. Stratford [a], Gunwoo Kim [ab], Rachel N. Kerber [a], Pieter M. Magusin [a], David A. Jefferson [a], Clare P. Grey [ab] *

a Chemistry Department, University of Cambridge, Lensfield Road, Cambridge, CB2 1EW, UK

b Cambridge Graphene Centre, University of Cambridge, Cambridge, CB3 0FA, UK

c Gonville and Caius College, Trinity Street, Cambridge, CB2 1TA, UK

d Diamond Light Source Ltd., Harwell Science and Innovation Campus, Didcot, OX11 0DE, UK

† Current address: Department of Material and Environmental Chemistry, Stockholm University, Svante Arrhenius Väg 16 C, Stockholm, 106 91, Sweden

*E-mail: cpg27@cam.ac.uk

**ABSTRACT:** We report the exfoliation of layered Na$_2$Ti$_3$O$_7$, a promising anode material for Na-ion batteries, and restacking using HNO$_3$ and NaOH to form H-[Ti$_3$O$_7$] and Na(x)-[Ti$_3$O$_7$] compositions, respectively. The materials were characterised by a range of techniques (SEM, TEM, solid-state NMR, XRD, PDF). Although the formation of aggregated nanoparticles is favoured under acidic restacking conditions, the use of basic conditions can lead to control over the adherence between the exfoliated layers. Pair distribution function (PDF) analysis confirms that the local TiO$_6$ connectivity of the pristine material is maintained. The lowest sodium-containing Na(1)-[Ti$_3$O$_7$] phase, which is the stable product upon Na$^+$ leaching after consecutive washing steps, displays the best performance among the compositions studied, affording a stable reversible capacity of about 200 mAh.g$^{-1}$ for 20 cycles at a C/20 rate. Washing removes the excess of 'free/reactive' Na$^+$, which otherwise forms inactive Na$_2$CO$_3$ in the insufficiently-washed compositions.

## Introduction

The discovery of atomically-thick graphene has recently aroused great interest in the properties and phenomena exhibited by two-dimensional (2D) materials, which in general can be considered as the exfoliation products of layered structures to form either single or few layers. The exfoliation of layered structures to individual or few nanosheets can be advantageous for applications requiring high surface activity, such as catalysis, electrochemistry and photoelectrochemistry. [1, 2]

In the extended family of layered inorganic structures, including metal oxides, metal chalcogenides (for example, reduced $TiS_2$, $MoS_2$ and $WS_2$), $LiCoO_2$ and others,[1, 2] interlayer counterions are often required to preserve/maintain the electroneutrality. This is advantageous for ion-exchange properties, but makes their exfoliation to generate individual layers more challenging than for graphite[1, 3, 4], because of the strong Coulombic forces that hold them together. [1]

In general, Na-ion batteries are considered as a lower-cost alternative to their Li-ion counterparts, which operate in a similar manner. This is in part due to the high natural abundance of sodium, along with the option to use aluminum current collectors on the anode side, instead of the more expensive copper used for Li-ion batteries. One of the key challenges for implementation of the Na-ion technology is related to discovering new anode materials since graphite, the anode of choice for Li-ion batteries, does not show electrochemical activity in Na-ion batteries.[5, 6] Hard (non-graphiteable) carbon can however reversibly intercalate $Na^+$ ions via a combined mechanism of $Na^+$ insertion between the nearly parallel layers and into nanopores but much of this storage takes place close to the sodium-plating voltage, raising potential safety concerns. [7, 8] Other potential anode materials are also being explored, [9, 10] among them, a variety of titanium-based structures, but these show low specific capacity in part due to the limited Na storage sites within the host structures. [9-11] Titanium-based structures are often preferred for various anode applications due to their lack of toxicity and relatively low cost compared to other transition metals, such as cobalt or manganese, and the redox activity of titanium being in the appropriate voltage window.[12]

Layered $Na_2Ti_3O_7$ is a well-known potential anode material for Na-ion batteries demonstrating good electrochemical properties at a low intercalation potential of around 0.3 V vs. $Na^+$/Na.[13, 14] It is built up of corrugated $(Ti_3O_7)^{2-}$ layers (formed by series of corner-sharing tri-octahedral ribbons) and $Na^+$ ions sitting in the interlayer space. The low intercalation potential of $Na_2Ti_3O_7$ is advantageous when coupled with a high-voltage cathode in a full cell to achieve a higher energy density. For example, $Na_2Ti_3O_7$ nanotubes have recently been tested as anode materials in full-cell batteries using the high-voltage cathode material $VOPO_4$ (V= 3.75 V vs. $Na^+$/Na), demonstrating promising electrochemical performance.[15] It is generally believed that reduction of $Na_2Ti_3O_7$ occurs via the reversible intercalation of two additional $Na^+$ ions in the structure, forming $Na_4Ti_3O_7$,[14] with 2/3 of the Ti(IV) ions being reduced to Ti(III).[13] Several studies have attempted to enhance the electrochemical capacity of this material by generating sodium vacancies within the crystal structure[16] and by fabrication of carefully controlled nanostructures. Recent advances in the field showing promising performance include bottom-up (BU) techniques for growing $Na_2Ti_3O_7$ based structures, such as growth of nanotube arrays[17], formation of a 3D spider-web architecture assembled by nanotubes[18], and growth on carbon-coated hollow spheres[19]. Promising results have also been reported when techniques favouring reduced particle size are used; these include crystalline rods prepared by microemulsion[20] and microspherical particles prepared by spray-drying[21]. Limited reports exist for preparing $Na_2Ti_3O_7$ via top-down (TD) facile techniques. It has been recently reported that exfoliation of bulk $Na_2Ti_3O_7$ can be achieved by ion exchange using alkyl ammonium ions. With some $Na^+$ ions remaining from the parent material, alongside propylamine ions added during the exfoliation process, a mixture of nanosheets and nanoplatelets was formed. This process led to an improvement in the charge storage kinetics and cyclability compared to the bulk material. [22]

In this study, we sought to entirely exfoliate $Na_2Ti_3O_7$ into nanosheets by complete removal of Na, along with the alkyl ammonium ions used during the process, followed by re-insertion of Na ions to restack the layers in a more disordered fashion. This was envisioned to create additional sites for $Na^+$ ion insertion, allowing for complete reduction of $Ti^{4+}$ to $Ti^{3+}$, thus increasing the capacity. [1-2] In addition, surface effects, as a result from exfoliation to thinner nanosheets, may favour further enhancement of the electrochemical performance in terms of the rate performance. We explored the effect of the exfoliation/restacking conditions on the structural chemistry, morphology and electrochemical properties of the phases obtained.

**Experimental**

Synthesis: The pristine layered $Na_2Ti_3O_7$ materials were made by solid-state synthesis, according to previous reports.[13] Stoichiometric amounts of $Na_2CO_3$ (Alfa Aesar, anhydrous, 99.5% purity) and $TiO_2$ (anatase, Sigma Aldrich, 99.8% purity) were ball-milled in isopropanol using a high-energy mill, and after evaporation of the solvent the mixture was sintered at 800 °C for 20 h in air, followed by grinding and a second sintering step under the same conditions. The same synthesis protocol was followed for the $Na_2Ti_6O_{13}$ and $Na_2TiO_3$ model compounds.

The protonated $H_2Ti_3O_7$ form was made by ion-exchange of pristine $Na_2Ti_3O_7$ via a solvothermal reaction in $HNO_3$ (aq) in an autoclave at 60 °C for 12 h. The material was isolated by filtration (Durapore PVDF membrane, pore size 0.22 μm, diam. 47 mm), washed with copious deionized (DI) water until neutral pH was achieved, and dried in an oven at 60 °C. The proton analogue $H_2Ti_3O_7$, was then gradually swollen by ion exchange in two steps of increasing size alkyl ammonium ions in aqueous solutions of amines (MA: methylamine and PA: propylamine). Each reaction was performed solvothermally at 120 °C for 48 h in an autoclave, with the concentrations used as reported previously.[23] The resulting colloidal suspensions were centrifuged at 6000 rpm for 5 min and the precipitates were washed and centrifuged for a further three times, until neutral pH. The $MA-Ti_3O_7$ and $PA-Ti_3O_7$ samples were then dried in an oven at 60 °C prior to further characterisation. The final titanate product of amine swelling ($PA-Ti_3O_7$) was then dispersed in water (0.1 g in 50 ml $H_2O$), followed by sonication in order to exfoliate to titanate layers/nanosheets. The non-exfoliated material was collected as the precipitate of centrifugation at 6000 rpm for 5 min. The supernatant suspension, containing the dispersed exfoliated nanosheets, was then restacked by mixing with $HNO_3$ or NaOH aqueous solutions, in 1:1 ratio by volume. After being left overnight to settle, the precipitates were easily separated by centrifugation (3000 rpm, for 3 min). In the case of restacking by NaOH, the washing procedure of the precipitate was carefully controlled, resulting in different amounts of inserted Na in the exfoliated/restacked samples. All the samples were dried, initially in air at 60 °C and finally at 100 °C in a vacuum oven.

Powder X-ray Diffraction (PXRD) measurements were performed in a reflection mode (Panalytical Empyrean diffractometer, with Cu K$\alpha_1$ radiation ($\lambda$=1.5406 Å).

Transmission electron microscopy (TEM) images were recorded at magnification between 40,000X and 800,000X in a JEOL JEM-3011 electron microscope operated at 300 kV, with objective lens characteristics Cs = 0.6 mm and Cc = 1.2 mm. With a $LaB_6$ emitter these produced an interpretable resolution limit of 0.17 nm and an absolute information limit of 0.14 nm. Scanning electron microscopic (SEM) images were recorded with a Hitachi S-5500 in lens field emission electron microscope.

Pair Distribution Function (PDF) analysis: Data of the samples, packed into kapton capillaries, were collected at the I15 beamline at Diamond Light Source, Didcot, UK. An X-ray beam of energy of 76 keV ($\lambda$ = 0.1631 Å) was used in conjunction with an amorphous silicon area detector (Perkin-Elmer). The sample geometry and the sample-to-detector distance were determined using a $CeO_2$ standard. The data were converted to intensity vs Q using the Data Analysis WorkbeNch (DAWN).[24] Standard corrections (background subtraction, Compton scattering, detector effects) were applied, and the data were Fourier transformed to obtain G(r) using the software PDFGetX2 using a $Q_{max}$ of 24 Å$^{-1}$.[25] Refinements against known $TiO_2$ phases were performed in PDFGui.[26] Refinements against single-layer models were performed in the Diffpy-CMI complex modeling framework.[27] The starting model was simulated from supercells of the $Na_2Ti_3O_7$ structure, where all atoms except for a single-layer in the middle of the unit cell were removed. The structure function was simulated from the Debye scattering equation [28] which was then Fourier transformed over a range of 1–24 Å$^{-1}$. The model was refined using a least-squares approach; unit cell parameters, *a*, *b*, *c* and *β*, an isotropic thermal parameter, $U_{iso}$, for each atomic species and a scale factor were allowed to refine.

Solid-state NMR: All solid-state $^1$H and $^{23}$Na NMR spectra were acquired on a 16.4 T

Bruker Avance III spectrometer using a 1.3 mm HX probe head. A rotor-synchronized Hahn-echo (for $^1$H) and a single-pulse sequence (for $^{23}$Na) were used to acquire magic-angle spinning (MAS) spectra with spinning frequencies of 55-60 kHz, recycle delays of 5 and 5-40 s (for $^1$H and $^{23}$Na, respectively), and radiofrequency (rf) field strengths of 125 and 140 kHz, respectively. $^1$H and $^{23}$Na shifts were externally referenced to solid adamantane at 1.87 ppm and solid NaCl at 7.21 ppm, respectively. Simulations of $^{23}$Na MAS powder spectra were performed with SIMPSON.[29]

First-principles calculations: DFT computations were performed using CASTEP with on-the-fly generated pseudopotentials. [30-32] For the electron–electron exchange and correlation interactions, the functional of Perdew, Burke and Ernzerhof (PBE)[33-36] was employed. Non-spin-polarized calculations were performed for the geometry optimization of the $Na_2Ti_3O_7$ system. The k-points mesh used was 3x3x3 with cutoff energy of 700 eV.

Compositional analysis: For C/H/N analysis, an Exeter analytical CE440 analyser was used. Typically, 1-2 mgs of sample were combusted at 950 °C in oxygen. For the ICP analysis, a Thermo scientific iCAP 7400 OES instrument was used. Samples were digested at 80 °C for four hours in 2 ml aqua regia. They were then made up to 50 ml using Millipore 18 M water. The ICP was calibrated using 10 ppm and 1 ppm standard samples.

Electrochemical testing was carried out in CR2032 coin cells vs. sodium metal, in a galvanostatic mode. The electrolyte used 1M $NaPF_6$ in propylene carbonate (PC, >99% Aldrich) and glass fibre (Whatman GF B 55) was used as a separator. The working electrode was prepared by mixing 70% active materials, with 30% carbon Super C-65 (Timcal) and 10% binder (PVdF-HPF copolymer) in NMP in a mortar and pestle. The mixed slurries were casted on copper foil using the doctor blade technique. A cell containing only carbon was also prepared for comparison reasons in a similar way, but the electrode slurry consisted of 70% carbon Super C-65 (Timcal) and 30% binder to ensure adequate adhesion to the copper foil and between particles. The laminates were vacuum dried at 100 °C for 12 h prior to punching and pressing the electrodes, followed by further drying at 100 °C under vacuum for 12 h. The cells were assembled in a glovebox under argon atmosphere ($O_2$ < 0.1 ppm and $H_2O$ < 0.1 ppm). The typical electrode loading was 20-25 mg.

## Results and Discussion

Pristine $Na_2Ti_3O_7$ was made by solid-state synthesis, as reported previously[13], and was characterized by PXRD (SI, Figure S1 and Table S1, also included in *Figure 1*B in red) and $^{23}$Na NMR (SI, Figure S2 and Table S2, spectrum also shown in *Figure 3*C). SEM shows that the pristine material comprises well-shaped nanorods of approximately 2.5 µm length and 0.1 µm thickness (*Figure 2*A).

The $Na_2Ti_3O_7$ exfoliation protocol used in this study is a modified version of previous reports for liquid exfoliation of titanates, niobates and titanoniobate nanosheets[23] and is illustrated in *Figure 1*A. It consists of the three successive processes of swelling, exfoliation and restacking. $Na_2Ti_3O_7$ swelling was accomplished via ion-exchange steps and the effect on the interlayer spacing (*d*) was monitored by PXRD (*Figure 1*B), based on the position of the 001 reflection. $Na_2Ti_3O_7$ was initially ion-exchanged to its protonated $H_2Ti_3O_7$ form. Complete ion exchange was achieved, in agreement with the literature [37], as confirmed by $^{23}$Na NMR (SI, Figure S2) and $^1$H NMR (SI, Figure S3).

Although the interlayer distance (*d*) in the proton analogue $H_2Ti_3O_7$ (*d* =7.87 Å) is smaller than in pristine $Na_2Ti_3O_7$ (*d* = 8.46 Å), the former is more reactive towards acid-base reactions and the layers may consequently be swollen by incorporation of protonated organic bases. This was done by ion exchange in two steps involving aqueous solutions of amines of increasing size (MA: methylamine, PA: propylamine) forming first MA-$Ti_3O_7$ and then PA-$Ti_3O_7$. Chemical analysis (SI, Table S3, C:N ratio) confirmed that the alkyl ammonium ions were intercalated as whole ions and did not decompose under the reaction conditions. The ion exchange was complete for the first amine exchange step, while the final product after intercalation of PA ($PA_{1.7\pm0.2}$)$Ti_3O_7$ was slightly amine deficient. $^1$H NMR of these materials (SI, Figure S3) confirm the presence and the stability of the alkyl ammonium ions; in the case of the PA-$Ti_3O_7$, a residual amount of protons remain in the structure bound to the $Ti_3O_7$ host. The yield of the exfoliation protocol strongly depends on the amount of sufficiently swollen titanate; the non-exfoliated content is removed at a later stage. The amine-driven swelling of the titanate framework results in general broadening of the reflections in the XRD patterns. The progressive shift of the 001 reflection position to lower angle indicates the increase of the interlayer spacing ($d$) upon intercalation of the alkyl ammonium ions ($d$= 10.13 Å and 13.55 Å for $H_2Ti_3O_7$-MA and $H_2Ti_3O_7$-PA, respectively). Upon enlargement of the interlayer distance along the $a$-axis, all the ($hk$0) reflections of $Na_2Ti_3O_7$ are expected to remain intact, with the 020 reflection at approximately 2θ (Cu K$_\alpha$) = 48 ° being the most intense. The (001) and (020) planes of pristine $Na_2Ti_3O_7$ are displayed in *Figure 1*F.

In order to exfoliate the titanate layers/nanosheets, the amine-swollen titanate product (PA-$Ti_3O_7$) was dispersed in water and then sonicated. After centrifugation to remove the precipitate containing non-exfoliated material, the supernatant containing the dispersed nanosheets was collected. The difference between the TEM of the exfoliated and the pristine material is striking. While the latter shows clearly-defined flakes and particles (SI, Figure S4), typical TEM images of the exfoliated (and dried) material (*Figure 1*C) are indicative of individual and multiple layers which have coalesced and collapsed as the solvent is removed, coming together in an extremely disordered arrangement with greatly reduced flake thickness. Arrays of layers can be seen in the high-contrast regions at the edges of the specimen (*Figure 1*D), but their arrangement is mostly irregular. In the regions where they form some sort of ordering, as for example at the very bottom of *Figure 1*D, no more than six layers with regular spacing are found. In these areas, the spacing between layers is in the order of 8 – 9 Å, similar to the regular layer spacing observed for the pristine material, indicating the structural similarity upon removal of the solvent. The exfoliated suspension was then restacked by NaOH; a representative SEM picture is displayed in *Figure 1*E showing layers stacked in a rather disordered fashion.

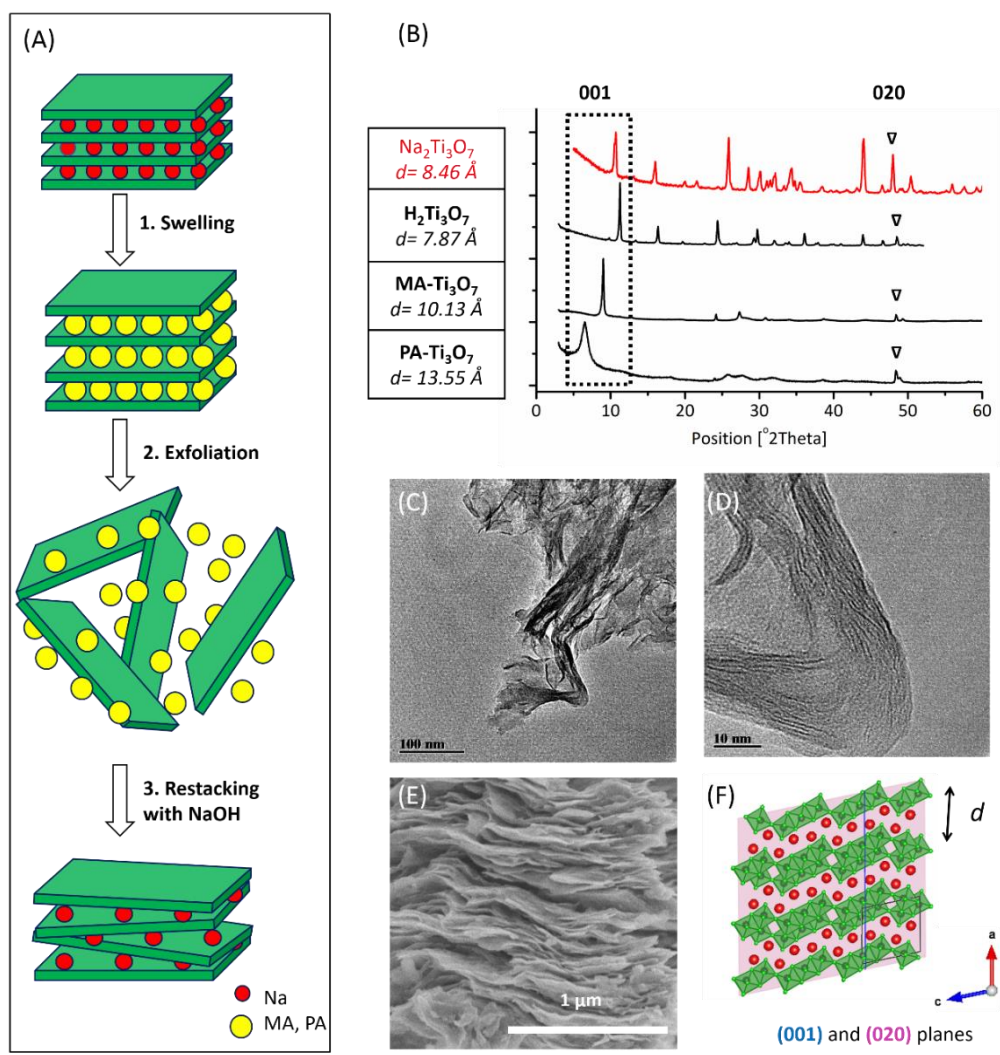

Figure 1: (A) Schematic of the exfoliation/restacking procedure, consisting of three consecutive steps: 1. swelling of pristine layered $Na_2Ti_3O_7$ by alkyl ammonium ions, by using aqueous solutions of amines of increasing size (MA: methylamine, PA: propylamine), 2. exfoliation by agitation and 3. restacking by NaOH; (B) Powder X-ray diffraction (PXRD) data monitoring the swelling, the dashed box highlighting the interlayer spacing ($d$), as derived from the 001 reflection. The pristine $Na_2Ti_3O_7$ was initially ion-exchanged to form the protonated $H_2Ti_3O_7$, which was then gradually swollen by ion exchange of the increasing size alkyl ammonium ions (MA, PA). The upturned triangle denotes the 020 reflection, which is retained upon swelling of the structure in the a-direction. (C, D) TEM images of exfoliated titanate nanosheet suspension, (E) representative SEM image of an exfoliated/ restacked specimen, (F) Crystal structure of pristine $Na_2Ti_3O_7$, highlighting the (001) and the (020) planes, the unit cell and the interlayer distance ($d$). Red spheres represent Na atoms and green $TiO_6$ octahedra form the layered framework.

In order to control and understand the restacking process, a careful investigation of the effect of different restacking conditions was made, with two series of experiments being performed. In the first method, $HNO_3$ (either 1 M or 2 M) was added to the nanosheet suspension, with the obtained white precipitate separated by centrifugation. For the second method, a large excess of NaOH (aq) was added to the nanosheet solution, followed by washing under different conditions.

The morphology of the products was determined by SEM (*Figure 2*). The nanorods observed for pristine ($Na_2Ti_3O_7$, *Figure 2*A) are still observed, but their shape is mostly irregular once the structure is swollen by the incorporation of alkyl ammonium ions (PA-$Ti_3O_7$, *Figure 2*B), which is the last step to exfoliation. Subsequent exfoliation and restacking changes drastically the morphology of the products, which varies according to the restacking conditions. Restacking by $HNO_3$ (H-[$Ti_3O_7$], *Figure 2*C) favours the formation of aggregated nanoparticles as also evidenced by the broad PXRD patterns, with the patterns fitting quite well with those for mainly $TiO_2$ anatase with some small amount of brookite (H(a)-[Ti$_3$O$_7$]) and rutile (H(r)-[Ti$_3$O$_7$]) polymorphs, for restacking by 1 M and 2 M HNO$_3$ respectively (SI, Figure S5). This is consistent with literature observations that different TiO$_2$ polymorphs can be formed hydrothermally from amorphous titania by varying the acidic conditions.[38] The fact that TiO$_2$ nanoparticles are the products of the exfoliation/(nominal) restacking process by HNO$_3$ suggests that the drying process used here results in structural transformation likely upon removal of structural water; this is reported to happen at temperatures above 80 °C.[39] However, restacking by NaOH (aq) results in the formation of stacked nanosheets. The stacking order seems to be strongly dependent on the washing procedure, with reduced order observed as the samples are further washed (*Figure 2*D, E and F).

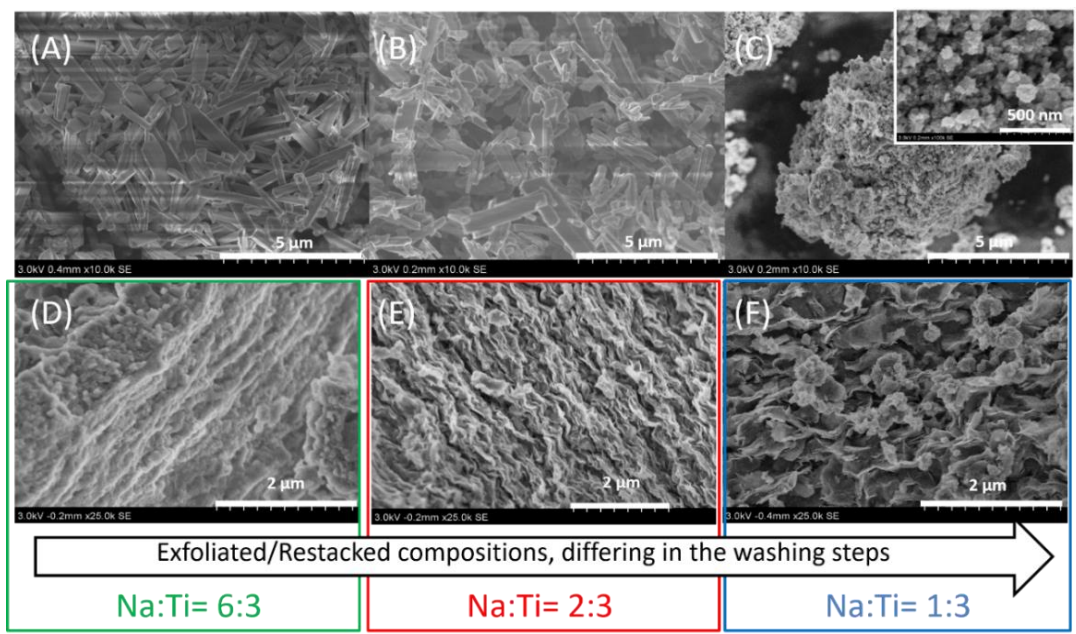

Figure 2: SEM images of (A) pristine Na$_2$Ti$_3$O$_7$, (B) PA-Ti$_3$O$_7$ (C) H(a)-[Ti$_3$O$_7$]; exfoliated/restacked composition by 1 M HNO$_3$, (D-E-F) exfoliated/ restacked compositions by NaOH, with general simplified formula Na(x)-[Ti$_3$O$_7$], showing decreased order of layers stacking with the number of washing steps. The Na:Ti ratio was measured by ICP/OES and was confirmed by converting the poorly crystalline phases to their crystalline analogues at high temperatures.

The composition of the nanosheets restacked by NaOH was first determined by ICP/OES (Table 1). The composition was also determined indirectly by converting the poorly crystalline Na(x)-[Ti$_3$O$_7$] phases at high temperatures into more well-defined crystalline phases; the Na:Ti ratio of the resulting phases as determined by PXRD (SI, Figures S6-S8) and confirmed by solid-state NMR (SI, Figure S10) provides a second estimate of Na content (assuming all the titanate material has been converted to crystalline material). The resulting compositions were expressed via the simplified formula Na(x)-[Ti$_3$O$_7$], so as to focus only on the Na(x) content (

Table 1).

**Table 1: ICP/OES results, PXRD indicating the major crystalline product identified after heating to 800 °C and composition Na(x)-[Ti$_3$O$_7$] determined from the composition of the crystalline phase.**

| Restacking and washing conditions | ICP (Na:Ti) | PXRD phase at 800 °C | Na(x)-[Ti$_3$O$_7$] |
|---|---|---|---|
| Large excess of NaOH (aq) | 1.89 | Na$_2$TiO$_3$ | Na(6)-[Ti$_3$O$_7$] |
| Mild washing conditions (x1) | 0.56 | Na$_2$Ti$_3$O$_7$ | Na(2)-[Ti$_3$O$_7$] |
| Washed thoroughly (x3), pH= 7 | 0.29 | Na$_2$Ti$_6$O$_{13}$ | Na(1)-[Ti$_3$O$_7$] |

The Na(x) content in the Na(x)-[Ti$_3$O$_7$] exfoliated/ restacked specimens was found to be approximately Na(6), Na(2) and Na(1) based on the PXRD analysis. Slightly lower Na contents were systematically determined by ICP/OES, which may be due to the techniques reported low sensitivity for Na, due to weak emission of alkali metal ions in general and interference issues for Na. [40] However, it may also reflect the presence of (amorphous) Na-deficient phase not seen by PXRD. The information obtained regarding the Na(x) content of the Na(x)-[Ti$_3$O$_7$] compositions demonstrate that washing induces leaching of Na$^+$ accommodated in the titanate exfoliated layers. Hence, control over washing can determine both the composition and the stacking order of the Na containing nanosheets. For easy reference, all of the exfoliated/restacked samples, and samples prepared with subsequent heat treatments and washings are listed in Table 2, along with the synthesized model compounds.

**Table 2: Samples prepared in this manuscript and details of synthesis conditions**

| SAMPLE | COMMENT | SYNTHESIS DETAILS |
|---|---|---|
| Na$_2$Ti$_3$O$_7$ | Pristine titanate | Made by solid state synthesis at 800 °C/40 h |
| Formed during intermediate steps of the exfoliation protocol | | |
| H$_2$Ti$_3$O$_7$ | Protonated titanate | Prepared by ion exchange of Na$_2$Ti$_3$O$_7$ in HCl (aq) solution |
| MA- Ti$_3$O$_7$ | Swollen H$_2$Ti$_3$O$_7$ by MA+ (MA+: CH$_3$NH$_3$+) | Prepared by ion exchange of H$_2$Ti$_3$O$_7$ in methylamine (MA) aqueous solution |
| PA- Ti$_3$O$_7$ | Swollen H$_2$Ti$_3$O$_7$ by PA+ (PA+: CH$_3$CH$_2$CH$_2$NH$_3$+) | Prepared by ion exchange of MA- Ti$_3$O$_7$ by propylamine (PA) aqueous solution |
| As made exfoliated/restacked compositions (all dried at 100 °C in vacuum oven) | | |
| H-[Ti$_3$O$_7$] | H(a)-[Ti$_3$O$_7$]: mainly anatase H(r)-[Ti$_3$O$_7$]: rutile | Exfoliated and treated with HNO$_3$ Drying leads to dehydration and to the formation of TiO$_2$ polymorphs |
| Na(6)-[Ti$_3$O$_7$] | Exfoliated/Restacked sample, with Na:Ti ratio=6:3 | Exfoliated and restacked with large excess of NaOH (aq) |
| Na(2)-[Ti$_3$O$_7$] | Exfoliated/Restacked sample, with Na:Ti ratio=2:3 | Exfoliated and then restacked material with large excess of NaOH (aq) and washed under mild conditions |
| Na(1)-[Ti$_3$O$_7$] | Exfoliated/Restacked sample, with Na:Ti ratio= 1:3 | Exfoliated and then restacked material with large excess of NaOH (aq) and washed thoroughly |
| Post-treatment conditions for exfoliated/restacked compositions Na(x)-[Ti$_3$O$_7$] | | |
| Na(x)-[Ti$_3$O$_7$]-400 | Dried at 400 °C for 12 h and then kept in vacuum oven at 100 °C | |
| Na(x)-[Ti$_3$O$_7$]-800 | Sintered at 800 °C for 8 h and then kept in vacuum oven at 100 °C | |
| Model compounds made for comparison reasons | | |
| Na$_2$Ti$_6$O$_{13}$ | Model compound, with Na:Ti ratio=1:3 | Solid state synthesis at 800 °C/ 40 h |
| Na$_2$TiO$_3$ | Model compound, with Na:Ti ratio=6:3 | Solid state synthesis at 800 °C/ 40 h |

The PXRD patterns of all exfoliated/restacked Na(x)-[Ti$_3$O$_7$] compositions are plotted together in *Figure 3*(A) and compared with those of crystalline NaOH.H$_2$O and Na$_2$CO$_3$ references, to exclude the presence of excess NaOH, as well as H$_2$O and CO$_2$ adsorption from the atmosphere. All the Na(x)-[Ti$_3$O$_7$] samples display very broad reflections, demonstrating their inherent disordered nature and absence of the 001 reflection at low angles, indicating that there is no significant ordered stacking of the layers. The PXRD pattern of the low-sodium phase Na(1)-[Ti$_3$O$_7$] only displays the 020 reflection at about 48 °2θ, which appears to be indicative of the structural coherence of the TiO$_6$ octahedral layered framework upon swelling (*Figure 1*B). This characteristic reflection is also observed in the patterns of the higher sodium containing phases Na(2)-[Ti$_3$O$_7$] and Na(6)-[Ti$_3$O$_7$], along with increasing amount of crystalline Na$_2$CO$_3$.

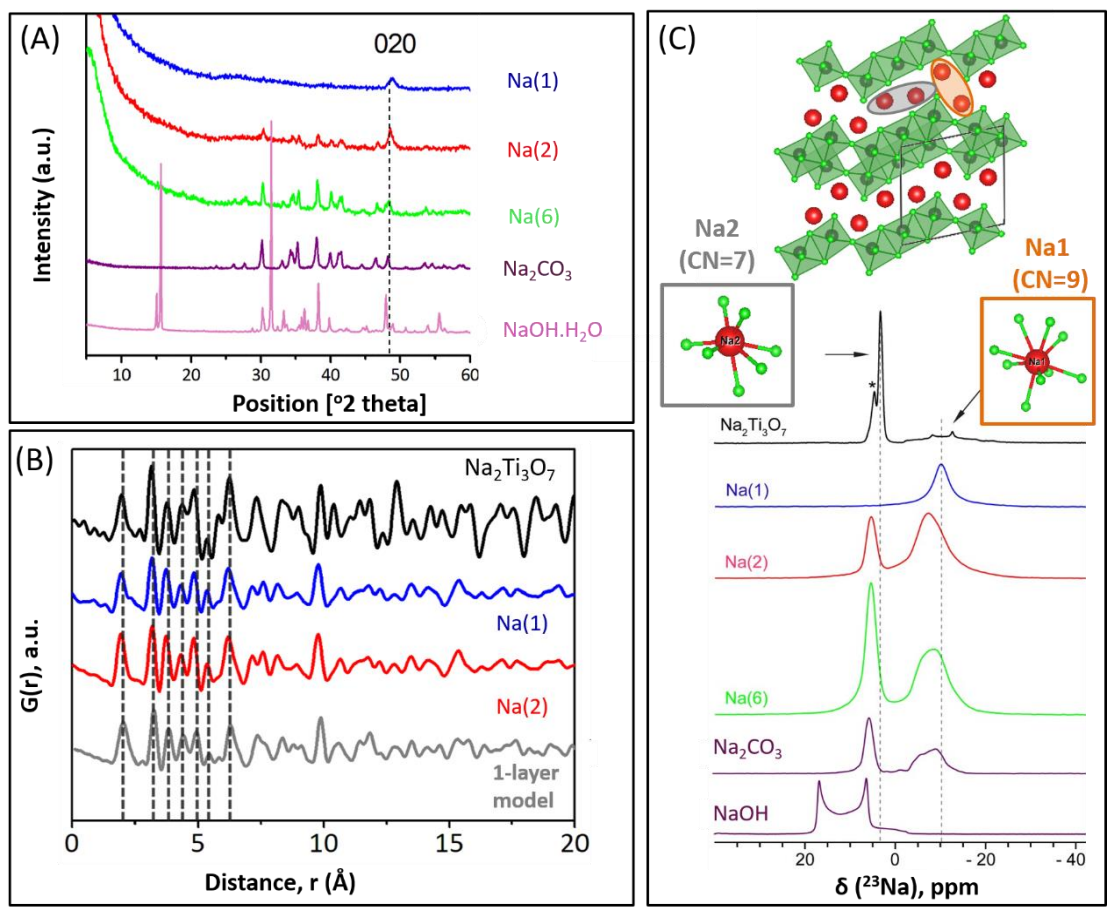

Figure 3: (A) PXRD patterns of all exfoliated/restacked Na(x)-[Ti$_3$O$_7$] compositions, including NaOH.H$_2$O and Na$_2$CO$_3$ reference compounds. (B) Pair Distribution Function (PDF) analysis data for pristine Na$_2$Ti$_3$O$_7$, Na(1) and Na(2) compositions and comparison with the PDF pattern simulated for a single layer. Dashed lines highlight the position of low-r peaks. (C) $^{23}$Na MAS NMR spectra acquired at 16.4 T for pristine Na$_2$Ti$_3$O$_7$, (highlighting the two different Na coordination environments in the structure) all exfoliated/restacked Na(x)-[Ti$_3$O$_7$] compositions, NaOH and Na$_2$CO$_3$ references.

*PDF analysis* In order to probe the local connectivity of the exfoliated materials, pair distribution functions (PDF) were extracted from total scattering data for Na(1)-[Ti$_3$O$_7$], Na(2)-[Ti$_3$O$_7$] and the pristine Na$_2$Ti$_3$O$_7$. These are shown in Figure 3B. Peaks are observed in the PDF beyond 50 Å (SI, Figure S11), implying that some long-range order exists and the material is not amorphous. Peak intensity dampens more quickly with increasing distance, *r*, in the exfoliated material compared to the pristine material; we assign this principally to the presence of disordered interlayer correlations (i.e. between Ti$_3$O$_7$ layers) resulting in a zero contribution to the total G(r). At longer interatomic distances, as the proportion of interlayer vs intralayer distances increases and, therefore, the total intensity of G(r) will decrease. In addition to this, the finite size/limited coherence of the nanosheets will also contribute to the drop of intensity with increasing distance. Below 5 Å, peaks in the experimental PDF for Na(1)-[Ti$_3$O$_7$] appear at very similar distances to the pristine Na$_2$Ti$_3$O$_7$ material (as shown by the dashed lines in Figure 3B), indicating that the immediate connectivity of the TiO$_6$ framework remains the same as the parent material.

Furthermore, the peak positions in the experimental data for Na(1)-[Ti$_3$O$_7$] are a poor match to all TiO$_2$ polymorphs (SI, Figure S12), and the residual factors obtained from least-squares refinements of TiO$_2$ polymorph structures against the experimental data for Na(1)-[Ti$_3$O$_7$] are very high (SI, Table S4). This confirms that the basic exfoliation/restacking conditions have not converted the material into other TiO$_2$ polymorphs, as was observed in acid conditions.

To further probe the extent to which the [Ti$_3$O$_7$] layers remain intact in the exfoliated material, a model of a single [Ti$_3$O$_7$] layer, approximately 50 x 50 Å in size was constructed (SI, Figure S13(a)). Only the contributions from Ti and O are considered in the model; no Na atoms were placed in the model because the $^{23}$Na NMR data indicate that a range of sodium environments exist (see below) and so are unlikely to contribute well-defined peaks to the PDF. Peak positions in the one-layer model show a good match to the experimental data, including the intense peak at 9.8 Å, which results from Ti-Ti interactions within the titanate layer. When this model was refined against experimental data for Na(1), a reasonable fit was obtained (SI, Figure S13 (a), R$_w$ = 0.39,), indicating that the structure of the exfoliated/restacked material is similar to that of a single layer of the Na$_2$Ti$_3$O$_7$ parent phase.

The fit remains imperfect; peaks in the residual of the refinement (G(r)$_{model}$ − G(r)$_{experiment}$ , grey line in SI, Figure S13 (a)) remain, indicating that some aspects of the structure are not captured by the current model and, therefore, the exfoliated/restacked materials structure shows some differences to Na$_2$Ti$_3$O$_7$-like Ti$_3$O$_7$ layers. There are several possible sources for these differences including: (a) additional broad contributions from the disordered sodium which are not accounted for in this model; (b) the presence of some inter-layer correlations, which although likely to be disordered, may lead to some changes in intensity; (c) additional structural complexity induced by defects in the exfoliated material, and; (d) a possible range of sheet sizes and connectivities. The fit to experimental data can be slightly improved by using a fragment with smaller dimensions, where some connectivity between the corner-sharing octahedra has been broken. Of the models tried (see SI, Section S4), a fragment with extended connectivity along the b-axis, but only a single set of corner sharing octahedra shows the best fit to the data (SI, Figure S13(b), R$_w$= 0.35). This implies that some breakdown/modification of the layers may have taken place during the exfoliation/restacking, either from sheet termination, or from defective areas within sheets. However, the calculated PDF for a single unit of six edge-sharing TiO$_6$ octahedra, the basic building block of the Ti$_3$O$_7$ layers shows no peaks beyond approximately 10 Å (SI, Figure S13 (d)). As interatomic correlations are observed well beyond this distance in the experimental PDF, this confirms that the sheets have not been broken down into very small units of the starting material. More detailed structural characterization will be the subject of another paper.

The principal difference between the two PDFs of the two restacked compositions is the intensity of the first peak, which is likely related to the presence of a low scattering component Na$_2$CO$_3$ in Na(2)-[Ti$_3$O$_7$] (as evidenced from XRD and NMR measurements) and/or difference in the amount of oxygen vacancies present.

*NMR Analysis* The $^{23}$Na NMR spectra of the three exfoliated/restacked (Na(x)-[Ti$_3$O$_7$]) compositions are shown in *Figure* 3C and compared with pristine $Na_2Ti_3O_7$ and NaOH and $Na_2CO_3$ references. The $^{23}Na$ NMR spectrum of $Na_2Ti_3O_7$ consists of two signals in agreement with previously reported data [41] and the Density Functional Theory (DFT) calculations performed in this work (SI, Figure S2). These are assigned to the two different Na coordination environments in the $Na_2Ti_3O_7$ layered structure,[42-44] the Na1 site at lower chemical shift (at around -10 ppm, nine-fold coordinated, coordination number (CN) = 9) with a significant second order quadrupolar broadening effect and the Na2 site (at approximately +3.2 ppm, seven-fold coordinated, CN=7). While the spectrum of the high sodium containing $Na(6)$-$[Ti_3O_7]$ phase is a good match with that of the reference $Na_2CO_3$ spectrum, in agreement with PXRD for high carbonate content, the low sodiated $Na(1)$-$[Ti_3O_7]$ phase only shows a single resonance with a shift at about -10 ppm. Although not displaying the characteristic second order quadropolar broadening, this resonance was labelled Na1' to reflect the similarity in shift position to that of the Na1 site (CN=9) in the pristine material. We do not necessarily assign this resonance to the same crystallographic site, but merely suggest that it has a similar CN. It is anticipated that there is some variation in the Na-O bond lengths in the disordered titanate network resulting in a distribution of slightly different chemical shifts and broader $^{23}Na$ NMR resonances compared to the pristine material. The intermediate sodium containing $Na(2)$-$[Ti_3O_7]$ phase has a spectrum that appears to correspond to a mixture of $Na_2CO_3$ and Na1', displaying the most intense shift for $Na_2CO_3$ at +5.5ppm and an asymmetric resonance at lower frequency (around -9 ppm), due to the presence of overlapping carbonate and Na1' signals.

All exfoliated/restacked compositions, display one $^1H$ NMR resonance with a shift at around +6 ppm (SI, Figure S14). This resonance could not be assigned to NaOH, and this signal is assigned to protons incorporated into the titanate network. $H_2Ti_3O_7$ gives rise to $^1H$ NMR shifts above +10 ppm (SI, Figure S3) and thus the proton signal is likely due to $H_2O/H_3O^+$ molecules/ions that are incorporated into the structure[45] (along with residual amine groups); the slightly displaced shift of +4.8 ppm is assigned to free water molecules. Spatial constraints associated with the order of nanosheet stacking (*Figure* 2D-F), which is enhanced with increasing the Na(x) content, could dictate the amount of water intercalated into the titanate network.

The absence of excess NaOH in the $Na(6)$-$[Ti_3O_7]$ and $Na(2)$-$[Ti_3O_7]$ phases is explained by the observation of $Na_2CO_3$, formed by reacting with $H_2O$ and $CO_2$ from the atmosphere; with $Na_2CO_3$ presumably crystallizing on the surface upon drying. A proposed mechanism is demonstrated in SI (Figure S15), which involves the formation of a $Na_2[Ti_3O_7]$* metastable phase (for x≥2), which transforms to the more stable $Na(1)$-$[Ti_3O_7]$ phase and liberates the excess $Na^+$. The excess $Na_2CO_3$ can cleanly be removed on washing, leaving $Na(1)$-$[Ti_3O_7]$ as the final stable phase. The Na atoms in $Na(1)$-$[Ti_3O_7]$ are accommodated in a high coordination site, additional coordination environments likely coming from bound water. (TGA and $^1H$ NMR indicate that this material contains approximately 2 $H_2O$ per $[Ti_3O_7]$ unit, SI, Figure S16). The titanate framework in the as-made $Na(1)$-$[Ti_3O_7]$ composition is preserved as shown from the PDF analysis. In order to maintain charge neutrality in this phase, oxygen vacancies may be present or more likely, the Na ions are coordinated by both water and $OH_3^+$ ions, the charge compensating protons being lost as water on heating. A transition to crystalline $Na_2Ti_6O_{13}$ ($Na_1Ti_3O_{6.5}$) is expected to occur once sintered at temperatures above 600°C [39] and is observed in this study as well when heated at 800°C (SI, Figure S8 and Figure S10 for $Na(1)$-$[Ti_3O_7]$) consistent to the cation ratio present in the exfoliated phase.

*Electrochemistry:* The exfoliated/restacked $H(x)$-$[Ti_3O_7]$ and $Na(x)$-$[Ti_3O_7]$ compositions were evaluated as anode materials for Na-ion batteries in the 0.005 -2.5 V range, at a C/20 rate (corresponding to the addition of 2 $Na^+$ in 20 hours). Preliminary data were also collected at C/10 and 1C for some of the compositions. The protocol initially adopted for pristine $Na_2Ti_3O_7$ with 30% carbon as a conductive additive was used for all samples to ensure consistency with the literature, and hand-grinding as opposed to mechanical milling was using to mix the reagents was used to avoid any structural rearrangement within the electrode materials. [13] Thus, the cycling behaviour of the cell containing only carbon was also investigated for comparison. The cycling data over 20 cycles of both H(a) and $H(r)$-$[Ti_3O_7]$ compositions are shown in *Figure* 4A, where closed and open symbols correspond to discharge and charge steps respectively. The first discharge capacity is notably high for both materials (≥1000

mAh.g$^{-1}$), indicating that a large amount of Na$^+$ atoms are consumed irreversibly by forming the solid electrolyte interphase (SEI) layer.

Although crystallising as different TiO$_2$ polymorphs, both specimens demonstrate similar electrochemical performance after the initial 10 cycles, both reaching about 150 mAh.g$^{-1}$ after 20 cycles, at C/20 rate. During the initial 10 cycles, the capacity values diverge with slightly higher values for H(r)-[Ti$_3$O$_7$] than H(a)-[Ti$_3$O$_7$] during discharge, and converge during charge steps. The capacity values for H(r)-[Ti$_3$O$_7$], with particle size distribution between 3-10 nm (*Figure* 2C and SI, Figure S5), are significantly improved when compared to commercial rutile TiO$_2$ (Sigma Aldrich #224227, particle size< 5 μm) measured at the same conditions for this study (*Figure* 4B). This demonstrates that nano-structuring TiO$_2$ has a significant effect on the electrochemical properties and it is the first report of nano-rutile being tested as a material for a Na-ion anode to the best of the authors' knowledge. [9, 10] Among the several TiO$_2$ polymorphs, the bronze-type TiO$_2$(B) has the most open crystal structure, followed by anatase, rutile and brookite in decreasing order, and is able to accommodate 1 Li+ per Ti giving a capacity of 335 mAh.g$^{-1}$ for Li ion batteries in both bulk and nanostructured forms[46]; much lower capacity values are observed for sodium ion batteries due to the larger size of the Na ions.[47] Nanosized TiO$_2$ typically shows poor cycling performance in Na-ion batteries; this is attributed to the formation of unstable SEI and side-reactions while cycling. [48-50]

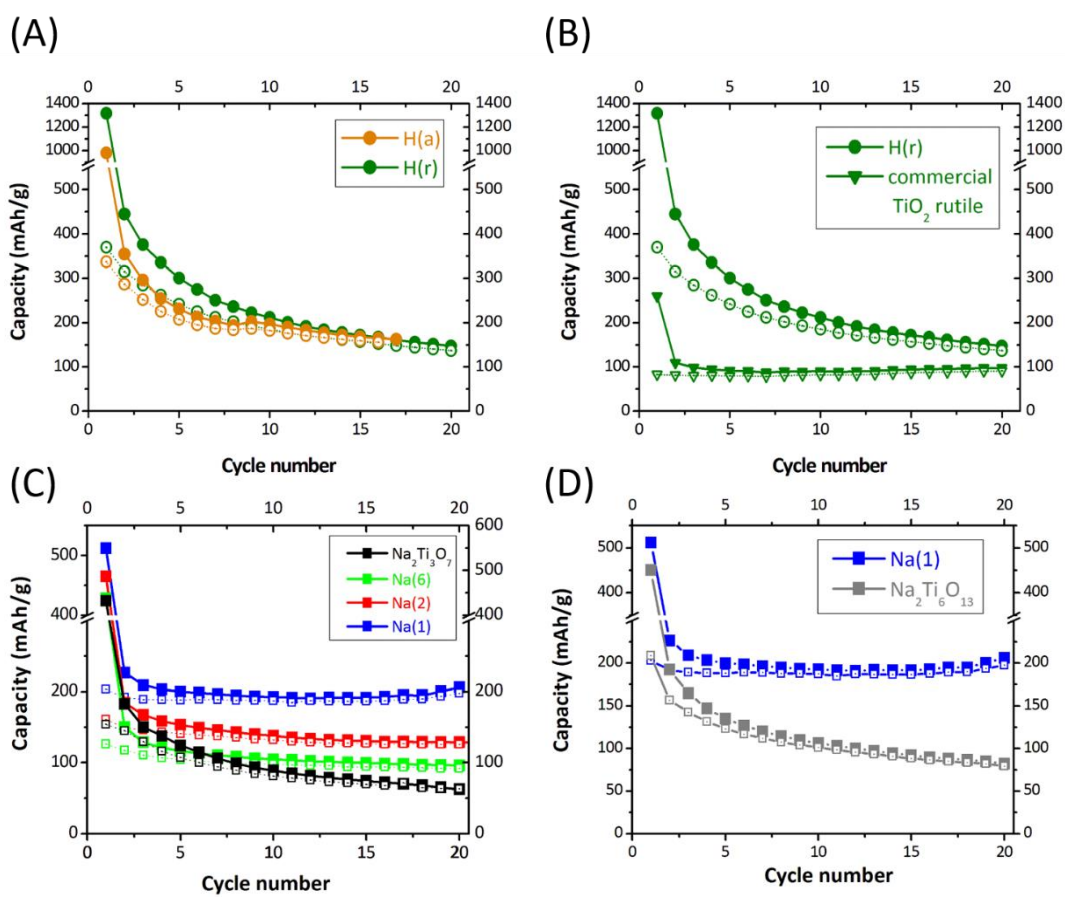

Figure 4: Cycling data over 20 cycles for (A) H(a) and H(r)-[Ti$_3$O$_7$] compositions; (B) comparison of H(r)-[Ti$_3$O$_7$] with commercial TiO$_2$ rutile; (C) all Na(x)-[Ti$_3$O$_7$] compositions and comparison with pristine Na$_2$Ti$_3$O$_7$. (D) Na(1)-[Ti$_3$O$_7$] and comparison with Na$_2$Ti$_6$O$_{13}$, which is the model compound with the same Na:Ti ratio. In all cases, the closed and open symbols correspond to discharge and charge steps respectively.

The cycling data over 20 cycles of Na(x)-[Ti$_3$O$_7$] compositions are shown in *Figure* 4C. The capacity increases with reduced Na(x) content, with Na(1)-[Ti$_3$O$_7$] displaying the best performance and attaining the highest stable capacity of about 200 mAh.g$^{-1}$ over 20 cycles. Despite the high first discharge irreversible capacity, similar to that obtained by other nanostructuring methods, [17-21] all the exfoliated/restacked Na(x)-[Ti$_3$O$_7$] compositions demonstrate improved cycling performance compared to pristine Na$_2$Ti$_3$O$_7$, which was measured under the same conditions and suffers from poor capacity retention[51]; this is a common problem for many potential materials used as anodes in Na-ion batteries. [10, 52] We note however that we have not attempted to optimize the electrode formulation. Further studies would be required to understand the effect that the first cycle losses have on long-term cyclability and in full cells.

The reduced capacity values observed for the Na(2)-[Ti$_3$O$_7$] and Na(6)-[Ti$_3$O$_7$] compositions compared to Na(1)-[Ti$_3$O$_7$] are ascribed to the increasing Na$_2$CO$_3$ content. However, the presence of Na$_2$CO$_3$ in these phases is not detrimental to the properties when compared to pristine and/or the model crystalline compounds, with the same Na:Ti ratio (*Figure* 4D for Na(1) and SI, Figure S17(A) and S17(B) for Na(6) and Na(2) respectively). This suggests that Na$_2$CO$_3$ is inactive and the improved capacity is related to the available space for Na intercalation and the amount of the active Na(1)-[Ti$_3$O$_7$] component.

Reduction of all the Ti(IV) content would be expected to occur via the intercalation of three sodium atoms and formation of Na$_5$Ti$_3$O$_7$ (C$_{theor}$= 267 mAh.g$^{-1}$) during cycling. In bulk Na$_2$Ti$_3$O$_7$, two sodium atoms are reversibly intercalated in the structure with concomitant reduction of 2/3 of the Ti(IV) to Ti(III) in Na$_4$Ti$_3$O$_7$ (C$_{theor}$= 177 mAh.g$^{-1}$).[13] The diverse values in the literature for pristine Na$_2$Ti$_3$O$_7$ depend largely on the formulation of the electrode, the additives and the choice of electrolyte, which has not as yet been standardized. [52] Further optimization studies would be required to evaluate the effect of electrode formulation and processing to the capacity values measured for Na(1)-[Ti$_3$O$_7$]. At this point, it should be noted that preliminary studies on intermediate compositions between the Na(2)-[Ti$_3$O$_7$] and Na(1)-[Ti$_3$O$_7$] end members, such as Na(ca.1.7), (SI, Figure S9 and Figure S10 for Na(ca.1.7)-[Ti$_3$O$_7$]) display higher capacities for initial cycles (e.g. about 240 mAh.g$^{-1}$ after 10 cycles) but these values are not stable and decrease to about 200 mAh.g$^{-1}$ after 20 cycles (SI, Figure S17(C)).

The galvanostatic curves of all exfoliated/ restacked Na(x)-[Ti$_3$O$_7$] compositions during the first cycle are shown in *Figure* 5A and compared with those of parent Na$_2$Ti$_3$O$_7$ and carbon cells. All the cells exhibit irreversible plateaus at about 0.6 – 1.0 V vs. Na$^+$/Na and at lower voltages (0.1-0.2V) during the first discharge. These are not observed during the first charge or in subsequent cycles, demonstrated here for the second cycle (*Figure* 5B) and the twentieth cycle (*Figure* 5C). The differential dQ/dV plots are shown in SI, Figure S18. These irreversible plateaus are believed to be due to reaction of Na$^+$ with the carbon additive and to the SEI formation from electrolyte decomposition.[13, 22, 51]

It should be noted that the existence of Na$_2$CO$_3$ in the Na(2)-[Ti$_3$O$_7$] and Na(6)-[Ti$_3$O$_7$] compositions may have an additional effect on the nature of the SEI forming during operation, due to its solubility in the electrolytes commonly used for sodium-ion batteries.[51] Moreover, the water molecules, whose existence was evidenced by $^1$H NMR (SI, Figure S14), likely participate to the processes occurring at the first discharge step. These intercalated water molecules were only removed upon further drying/treatment at 400 °C (SI, Figure S16), which is likely to have an effect to the nanosheet stacking. The capacity values of Na(x)-[Ti$_3$O$_7$]-400 compositions were reduced by about 100 mAh.g$^{-1}$ during the first discharge step compared to as-made samples, while there was no significant difference for the subsequent cycles (SI, Figure S19). Although the incorporation of water molecules would be expected to influence the electrochemical performance while cycling by continuously reacting with the electrolyte, this was not observed for these samples under the conditions tested. It is however anticipated that the existence of intercalated water molecules might affect further high rate capability studies.

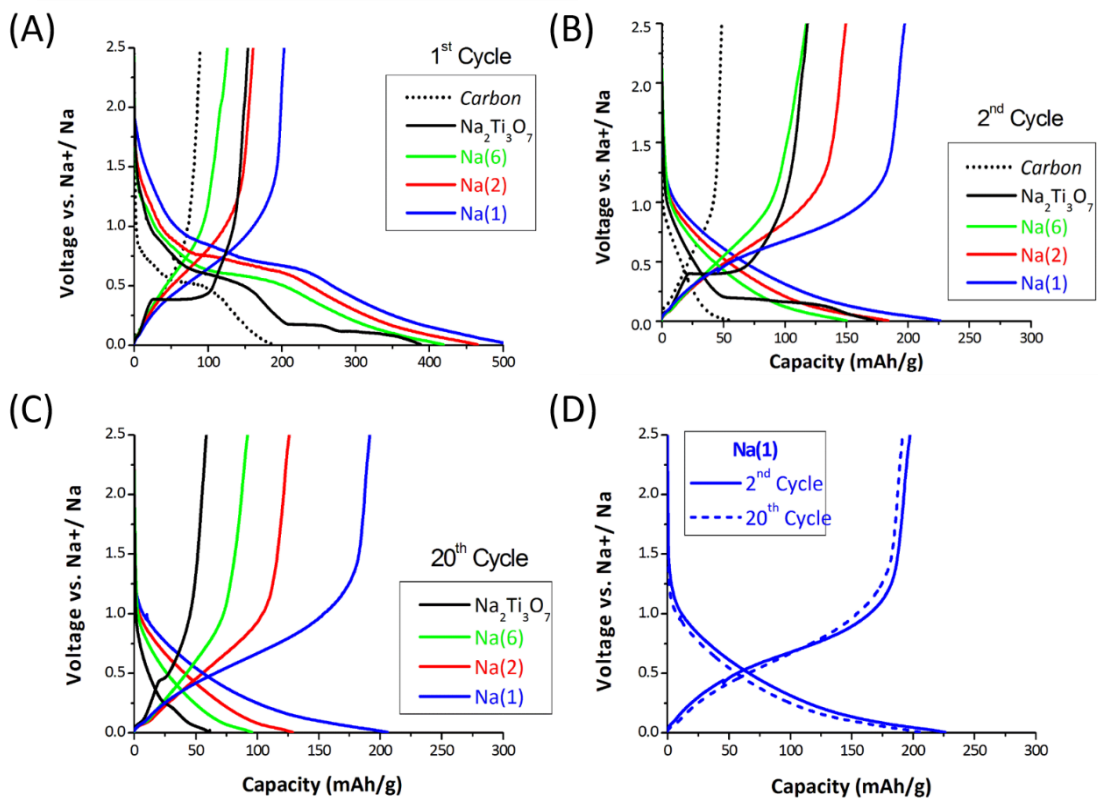

Figure 5: Galvanostatic charge/discharge curves for all exfoliated/restacked Na(x)-[Ti$_3$O$_7$] compositions and comparison with pristine Na$_2$Ti$_3$O$_7$ and carbon cells, measured at the same conditions for (A) the 1st, (B) the 2nd and (C) the 20th cycles. (D) Direct comparison of the 2nd and 20th cycles for Na(1)-[Ti$_3$O$_7$].

After the formation of SEI during the first discharge step, the electrochemical processes for the pristine Na$_2$Ti$_3$O$_7$ occur at distinct active sites in the bulk favoring a two-phase reaction, accompanied by flat voltage plateaus at around 0.2 and 0.4 V vs. Na$^+$/Na during discharge (sodiation) and charge (de-sodiation) steps, respectively[13] (*Figure* 5B and *Figure* 5C for the 2$^{nd}$ and 20$^{th}$ cycles, respectively, dQ/dV plots in SI, Figure S18). The additional small plateau at about 0.1 V vs. Na$^+$/Na, observed for pristine and all exfoliated/restacked compositions, is believed to be related with the carbon additive (SI, Figure S18(D)). On the contrary, all the cells containing the exfoliated/restacked Na(x)-[Ti$_3$O$_7$] compositions display sloping discharge/ charge potential profiles (shown separately for Na(1) in *Figure* 5D) compared to the pristine material, with a very broad plateau at average voltage about 0.6 V vs. Na$^+$/Na, most intense for Na(1)-[Ti$_3$O$_7$] composition (also observed at dQ/dV plots, SI, Figure S18). This plateau is likely to be related to the existence of the Na intercalation sites (as evidenced by $^{23}$Na NMR, *Figure* 3C) and is not observed for the H-[Ti$_3$O$_7$] compositions (SI, Figure S20).

In general, sloping potential profiles are indicative of single-phase behaviour, where the intercalation/deintercalation processes occur at non-equivalent sites with different energies, in part due to the disorder of these systems and the large exposed surfaces of reduced particle-size materials. [53, 54] The overall surface area of the exfoliated/restacked materials in this study, is expected to be higher than for the pristine crystalline Na$_2$Ti$_3$O$_7$ (BET = 1.3 m$^2$.g$^{-1}$); reliable BET measurements would require synthesis on a larger scale. In the general case of nano-sized materials, the intercalation/ deintercalation processes are accompanied by shortened diffusion/ transport distances, which is often advantageous for the overall specific capacity.[54] For the restacked Na(x)-[Ti$_3$O$_7$] nanosheets, the existence of Na intercalation sites is likely related to the observed improved capacity retention. Although higher rate data were not collected for Na(1)-[Ti$_3$O$_7$], preliminary data for Na(2)-[Ti$_3$O$_7$] indicate that the capacity of about 120 mAh.g$^{-1}$ remains stable when cycled at C/20 and 1C for 40 cycles (SI, Figure S20(F)). Clearly, there is further scope for optimization and high rate studies.

## CONCLUSIONS

In summary, we report the exfoliation and subsequent restacking of $Na_2Ti_3O_7$ using $HNO_3$ and $NaOH$ to form $H\text{-}[Ti_3O_7]$ and $Na(x)\text{-}[Ti_3O_7]$ compositions, respectively; these were tested as potential Na-ion anode materials. The formation of aggregated $TiO_2$ (rutile and anatase) nanoparticles is favoured upon drying of the $H\text{-}[Ti_3O_7]$ compositions, resulting in high capacity materials, but which suffer from poor capacity retention. In the case of the $Na(x)\text{-}[Ti_3O_7]$ compositions, the local connectivity of the titanate framework was retained. Control over the composition can be achieved by washing, which induces gradual Na leaching from the titanate network resulting in reduced degrees of adherence between the exfoliated layers. The final composition is $Na(1)\text{-}[Ti_3O_7]$, washing removing the excess of "free/reactive" $Na^+$, which otherwise forms inactive $Na_2CO_3$ in the insufficiently washed compositions. All the exfoliated/restacked $Na(x)\text{-}[Ti_3O_7]$ compositions show significantly improved capacity retention compared to the pristine $Na_2Ti_3O_7$ and/or the crystalline compounds with the same Na:Ti ratio. This is likely related to enhanced kinetics due to nanosize effects and the formation of a more open titanate framework. The optimal electrochemical performance among the series of exfoliated/restacked $Na(x)\text{-}[Ti_3O_7]$ compositions studied is demonstrated for the $Na(1)\text{-}[Ti_3O_7]$ phase, where Na atoms likely occupy a range of slightly different Na intercalation sites, $Na(1)\text{-}[Ti_3O_7]$ displaying a stable capacity of about 200 mAh.g$^{-1}$ after 20 cycles at C/20 rate. In-situ and/or post-cycling studies would be required to conclude whether the intercalation of additional sodium atoms compared to pristine $Na_2Ti_3O_7$ is dominating over surface factors.

## SUPPORTING INFORMATION

Further details about the characterization of materials (pristine, intermediates and products) by chemical and NMR measurements, PDF, XRD and TGA analysis, DFT calculations and TEM images, as well as supplementary electrochemical data and analysis, including Tables S1-S4 and Figures S1-S20.

## NOTES

Additional data related to this publication are available at the Cambridge data repository: https://doi.org/10.17863/CAM.18615

## ACKNOWLEDGMENTS

We thank Diamond Light Source for the provision of beam- time and access to beamline I15 (EE-14177) that contributed to the results presented here and the beamline scientists Dr. Phil Chater and Dr. Annette Kleppe. The authors thank Dr Wanjing Yu, Dr Elizabeth Castillo-Martinez and members of Cambridge Graphene Centre for useful discussions, Dr. Tao Liu for help with the SEM images and Dr. Abdul R.O. Raji for the BET measurements. M.A.T. acknowledges funding from the Engineering and Physical Sciences Research Council (EPSRC) of the United Kingdom (Grant no. EP/L019469 and EP/K01711X/1) and the EU Graphene Flagship under contract no. 604391. M.A.T, A.J.P, J.M.S and R.N.K. acknowledge funding from the United States Department of Energy (DOE, funder reference: 7057154). P.K.A. acknowledges a Junior Research Fellowship from Gonville and Caius College and an Oppenheimer Fellowship from the University of Cambridge. This project has also received funding (R.N.K) from EPSRC (Grant no. EP/K030132/1) and from the European Union's Horizon 2020 research and innovation programme under grant agreement No. 696656–GrapheneCore1 (G.K.).

TOC

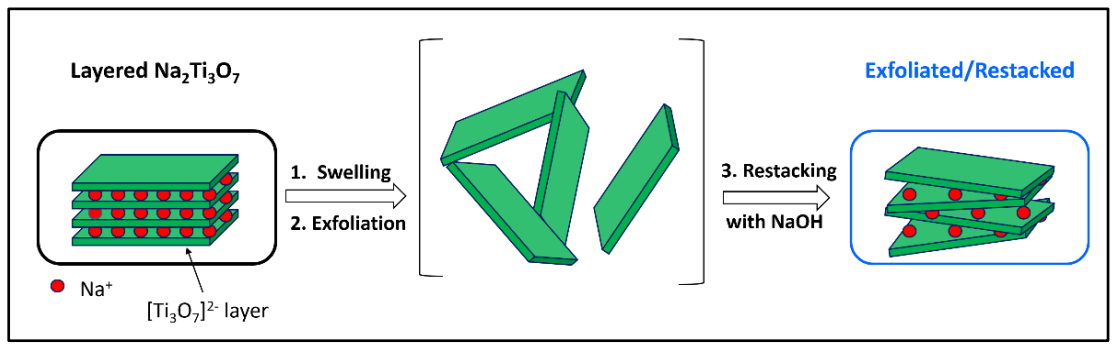